# Coherence in a simple network: Implication for brain function★

Zhen Ye

*Department of Physics and Center for Complex Systems, National Central University, Chungli, Taiwan 32054*

(May 28, 2000)

In a many body system, constituents interact with each other, forming a recursive pattern of interaction and giving rise to many interesting phenomena. Based upon concepts of the modern many body theory, a model for a generic many body system is developed. A novel approach is proposed to investigate the general features in such a system. An interesting phase transition in the system is found. Possible link to brain dynamics is discussed. It is shown how some of the basic brain processes, such as learning and memory, find therein a natural explanation.

## I. INTRODUCTION

A major task in physics research is to search for general principles behind various nature processes. One of the greatest advances in the last century is understanding of how a many body system manifests and behaves, leading to many important phenomena including the multiple scattering of both quantum and classical waves, the quantum Hall effect, the transition between super-conducting and normal metal states and so on. It has become a daily experience that macroscopic objects arise from interaction between individuals in a many body system which may be of microscopic scale.

In this way, we are led to a fundamental question: are there general features which can be said about the manifestation of many body interactions. Addressing this question, many profound concepts have been developed and matured over the past half century. To name a few, this includes the phase transition, invariance or symmetry, and symmetry breaking; these concepts fabricate the quilt of the modern many body theory [1]. An intricate inter-connection among these is stated by the Goldstone theorem. A symmetry breaking will result in not only a phase transition, but also the appearance of some sort of long range order, or synchronization, or coherence, depending on what subject is under scrutiny. For example, the phase transition between the liquid and solid states is a result of breaking the continuous space invariance. Once the symmetry breaks, phonon modes appear to reflect the collective behavior of the system. Although they can be systematically obtained in the realm of the quantum field theory, for instance, from the Ward-Takahashi relation, these phonon modes are not of quantum origin. This observation leads us to believe that the long-range macroscopic coherence or order does not necessarily require a quantum cause.

A natural question may be whether the many body theory may be applied to the problem of brain, a most complicated classical hierarchy kingdom. Physical research on brain function is not merely out of the scientific curiosity, but rather the problem is one of the most basic research subjects and has long been an outstanding inquiry from various scientific communities.

Motivated by our recent work on wave localization [2], in this communication, we will formulate a general model for interacting many body systems. It will be clear that this model can represent many important physical situations. Then a new approach will be proposed to study the model. Numerical examples will be presented, followed by a discussion on their implications for brain function.

## II. THE MODEL

Without compromising generality and for the sake of simplicity, we start with considering a system consisting of $N$ electric dipoles. These dipoles can be either *randomly* or *regularly* distributed in a space which can be either one, two, or three dimensional. The former forms random arrays of objects, while the latter would form a crystal lattice allowing band structures to appear. The dipoles interact with each other through transmitting electrical waves when each dipole oscillates. A conceptual layout is presented in Fig. 1

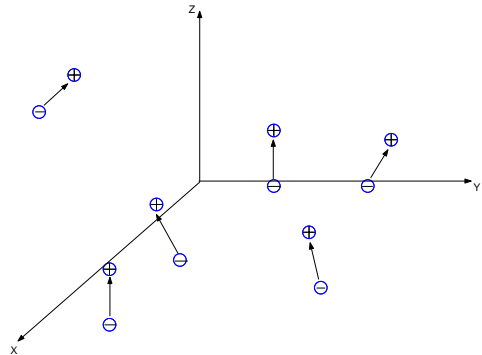

FIG. 1. A conceptual layout of a many dipole system

The equation of motion for any dipole can be easily deduced from the electro-magnetic theory [3]. Neglecting terms higher than the second order, for the $i$-th dipole, for example, the equation of motion for the dipole oscillation is

$$\frac{d^2 Q_i}{dt^2} + \omega_i^2 Q_i + \gamma_i \frac{dQ_i}{dt} = F_i(t)$$
$$+ \sum_{j=1, j\neq i}^{N} C_{ij} \left(1 - \frac{1}{2} Q_i \hat{z}_i \cdot \nabla\right) \frac{D^{ij}}{Dt^2} \frac{Q_j(t - |\vec{r}_i - \vec{r}_j|/c)}{|\vec{r}_i - \vec{r}_j|},$$

where

$$\frac{D^{ij}}{Dt^2} = [(\hat{r}_{ij} \cdot \hat{z}_i)(\hat{r}_{ij} \cdot \hat{z}_j) - 1]\frac{d^2}{dt^2}.$$

In these, $Q_i$ is the oscillation displacement with $\omega_i$ being the natural frequency and $\gamma_i$ being the damping factor, $\hat{z}_i$ is the unit vector along the dipole direction of the $i$-th dipole, $\hat{r}_{ij}$ is the unit vector pointing from the $j$-th dipole to the $i$-th dipole, the coupling constant $C_{ij} = \frac{-q_i q_j \mu_0}{2\pi m_i}$. In the present paper we ignore the rotational motion of the dipoles, which can be added with ease.

The second term on the RHS of Eq. (1) denotes the nonlinear interaction. For the first step analysis, this effect may be dropped. Equation (1) can formally be generalized as

$$\mathcal{L}_i(\partial)\psi_i = F_i + \sum_{j=1, j\neq i} \hat{T}_{ij}[\psi_j], \tag{2}$$

where $\psi_i$ is the physical quantity under consideration; in the dipole case $\psi_i$ is the displacement, $F_i$ refers to the external stimuli at the $i$-th entity, $\hat{T}_{ij}$ is an operator transferring interaction between the $i$-th and $j$ component which usually can be written in terms of a Bosonic propagator. The operator $\mathcal{L}(\partial)$ is called divisor obtained when the many body interaction is not present [1]. So far, in nature there are two known types of divisors. Type I usually refers to the boson like systems and is written as

$$\text{Type I}: \quad \mathcal{L}(\partial) = \frac{1}{c^2}\frac{\partial^2}{\partial t^2} - \omega^2(\nabla).$$

Type II divisor usually refers to the fermion like systems and is written as

$$\text{Type II}: \quad \mathcal{L}(\partial) = i\hbar\frac{\partial}{\partial t} - \epsilon(\nabla).$$

In the above $\omega(\nabla)$ and $\epsilon(\nabla)$ refer to the operators that give the energy spectra in absence of interaction, i. e. the bare energy.

Equation (2) accommodates a variety of problems of great interest.

**Solid lattices.** The vibration of the lattice formed by atoms is described by

$$(\Phi_{ll} - \omega_l^2)u_l + \sum_{l'\neq l}\Phi_{ll'}u_{l'} = 0, \tag{3}$$

where $\Phi_{ij}$ is the force term, $u_l$ is the displacement, and $\omega_l$ is the natural frequency. Clearly this equation resembles Eqs. (1) and (2).

**Spin arrays.** From the Hamiltonian

$$H = -\frac{1}{2}\sum_{ll'} J_{ll'} \vec{S}_l \cdot \vec{S}_{l'},$$

we define $S_l^\pm = S_l^z \pm i S_l^y$, then

$$\left[\left\{2S\sum_{l'\neq l}J_\parallel(ll')\right\} - \hbar\omega\right]S_l^- - 2S\sum_{l'\neq l}J_\perp(ll')S_{l'}^- = 0 \tag{4}$$

where $S$ is the total spin, and $J_{\parallel,\perp}$ are the couple constants. This is also in the form of Eq. (2).

**Electronic systems.** From the Shrödinger equation

$$\left[-\frac{\hbar^2}{2m}\nabla^2 + V(\vec{r})\right]\psi(\vec{r}) = E\psi(\vec{r})$$

with

$$V(\vec{r}) = \sum_l V_l(\vec{r} - \vec{R}_l).$$

In terms of LCAO (Local Combination of Atomic Orbits), we obtain

$$\psi(\vec{r}) = \sum_l u_l^\alpha(\vec{r} - \vec{R}_l),$$

then

$$(\epsilon_l^\alpha - \epsilon)u_l^\alpha + \sum_{l\neq l'}V_{ll'}^{\alpha\beta}u_{l'}^\beta = 0. \tag{5}$$

This is the Type II version of Eq. (2).

**Multiple scattering of classical waves.** Multiple scattering of waves is established by an infinite recursive pattern of rescattering between scatterers. The process can be solved from a set of coupled equations

$$p_s(\vec{r}, \vec{r}_i) = \Pi_i\left(p_0(\vec{r}_i) + \sum_{j\neq i}p_s(\vec{r}_i, \vec{r}_j)\right)G^{(d)}(\vec{r}, \vec{r}_i), \tag{6}$$

where $G^d$ is the propagator in $d$ dimension. This is a version of Type I equation in the momentum space.

In summary, Eq. (2) represents a wide range of many body problems. It reflects a basic many body interaction picture. The interaction among the constituents is mediated by waves. This is common in nature. We would like to explore some new features embedded in Eq. (2). We consider wave propagation in a Type I system. The incident wave will be interfered by radiated waves from each entity - the radiated waves are a response to the incident wave. The energy flow is $\vec{J} \sim \text{Re}[u^\star(-i\nabla)u]$. Writing the field as $Ae^{i\theta}$, the current becomes $A^2\nabla\theta$, a version of the Meissner equation. This hints at that when $\theta$ is constant while $A \neq 0$, the flow stops and energy will be localized in space. This phase transition is condensation of modes in the real space, in contrast to the Cooper pair condensation in the momentum space in the superconductivity. This is useful as it implies that energy can be stored in certain spatial domians; as will be clear, it links to some basic features of brain function.

## III. RESULTS AND IMPLICATIONS FOR BRAIN FUNCTION

We solve Eq. (2) for Type I situations in the frequency domain. There are $N$ bodies in the system, referring to Fig. 1 , and they are excited by a point source located in the middle of these bodies. To further simplify our discussion, we assume that all elements in the system are identical and the interaction is isotropic; the study of anisotropic cases is possible but comuptationally expensive and some examples have been documented elsewhere [4].

We write
$$\psi_i = A_i e^{i\theta_i}, \qquad (7)$$
where we dropped the common time factor $e^{-i\omega t}$. The total wave in the space is the summation of the direct wave from the source and the radiated waves from all the bodies. For each phase $\theta_i$, we define a phase vector such that
$$\vec{v}_i = \cos\theta_i \vec{e}_x + \sin\theta_i \vec{e}_y. \qquad (8)$$

This phase vector can be placed on the respective entity and put on a 2D plane spanned by $\vec{e}_x, \vec{e}_y$. The amplitude $A_i$ refers to the energy distributed among the entities.

Numerical computation has been carried out for 1D, 2D and 3D situations respectively. The results are summarized as follows. (1) In the 2 and 3 dimensional cases, for sufficient large coupling strength $C$ and enough density of bodies but small damping factor, the energy is trapped inside the system for a range of frequencies located above the nature frequency. More explicitly, at a frequency in that range, the transmitted wave from the exciting source is localized inside the system. There is no energy flow, as supported by the appearance of a global coherence in the phase vectors defined above, i. e. all phase vectors point to the same direction and phase $\theta$ is constant. (2) When the damping increases, transferring more and more energy to other forms, the coherence phenomenon starts to disappear gradually. (3) Decreasing the density of the constituents also allows the energy to be released from the system, meanwhile the coherence disappear. Another effect of decreasing density is to narrow the trapping frequency range. (4) Outside the trapping frequency range, no energy can be stored in the system. Therefore adjusting the frequency will crucially tune whether the system can store energy. (5) For 1D cases, energy is trapped near the source for all frequencies with any given amount of random placement of the entities. (6) Last, perhaps most importantly, all these features are not sensitive to any particular distribution of the constituents. In other words, these features remain unchanged as the configuration of the system is varied.

These features are illustrated by a 2D situation shown in Fig. 2 Here is shown that no energy is stored for $f/f_0 = 0.88235$ and $5.8824$, but the energy becomes trapped for $f/f_0 = 1.9608$. Clearly there are phase transitions from $f/f_0 = 0.88235$ to $1.9608$, and from $1.9608$ to $5.8824$. We stress that these features do not require quantum origin. Here $f_0$ refers to the nature frequency.

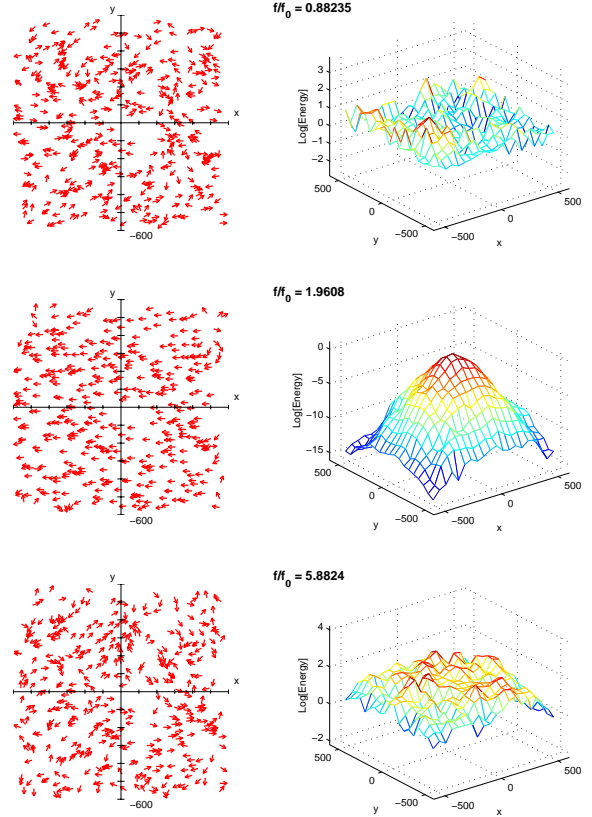

FIG. 2. Energy distribution and diagrams for the phase vectors in a 2D random configuration. Right: Energy distribution; the geometrical factor has been scaled out. Left: the phase diagrams for the phase vectors.

Now we are ready to discuss the linkage to brain function. Although brain may be the most complex system in the nature, still several general observations (dogmas) can be stated. First, brain is a system of many body. Any single neuron should not be significant for the whole brain, but rather the patterns of activity of clusters are important. As early as in 1967, Ricclardi and Umezawa first proposed to explain the brain dynamics in the context of many body theory [5]. Their work has stimulated and continues to inspire much research along the line. A special area in brain research has emerged and is known as quantum brain dynamics [6]. Second, experiments have indicated that there are coherence phenomena in brain function. Neurons or neuron clusters at different spatial points act collectively, performing a specific task. Such coherence is also observed for microtubles in nerve cells [7] and other biological systems [8]. To effectively performing long range collective tasks requires the communication to be mediated by waves and the brain system to be boson-like. Unlike the previous research, it is the present author's view that the brain function is not of quantum origin. Third, brain not only accumulates but stores energy, corresponding to learning and memory processes respectively. As memories in a healthy brain are flashed in a regular pattern, it is reasonable to be-lieve that the memory as a form of energy is stored in a coherent manner. Frölich was the first to link energy storage with long range coherence in certain biological structures [8]. Fourth, the brain function can resist reasonable damages which may cause the brain structure to dislocate to a rational extent. This implies that the brain function can be executed in relatively disordered environments. Lastly, brain can age, losing memory and becoming dying.

All these important features of brain can be well explained in the realm of the above generic model. Although it is not possible to discern the exact entities that carry out brain function at this stage, a few possibilities have been mentioned in the literature, such as neurons or corticons, proteins in microtubules and so on. Even water molecules are also mentioned as a core role in memorizing and storing. Putting details aside, what is important is only a *quantity* somehow related to the activity of brain. Actually, it is known that neurons do not seem to be the only fundamental units of the brain; or, at least, together with them some other elements, as for instance *glia celles*, may play an important role [9]. No matter what they exactly are, the necessity for stability requires the constituents to follow Type I equation. The communication between entities should be mediated by waves to ensure effective information exchange. Based on the above results, we argue that the brain consists of many subsets of Type I entities, each subset has different natural frequencies, and accomplishes different duties. Different subsets will store specific information depending on the spectra of external stimuli. Once the stimuli fits into a trapping range of a specific subset, the energy of characters of the stimuli will be stored or imprinted into that subset in a coherent manner, as indicated by Fig. 2. In other words, different subsets response to respective stimuli and store energy accordingly, corresponding to learning and memory processes. When the density of entities is not high enough or there is no enough active entities, however, the energy cannot be stored for long time. This reveals as the situation of short memory; this may be seen, for example, in the early development of children's brains. Long term memory requires sufficient amount of active entities and reasonable supply of stimuli. Moreover, as mentioned above, the feature of long range coherence and energy storage remains unchanged as the entity configuration varies; here we exclude the extreme case that all entities collapse into each other. This would explain the brain's resistance to reasonable damages. Damages may reduce the active entities and alter the structure of the system. Either varying the number of active entities or the natural frequency of entities or both produces the different states of the brain and controls the energy release. Memories may also be lost and disoriented when the the damping effect becomes prominent. The possible contributions to the damping effect include that aging effect and diseases.

## IV. CONCLUDING REMARKS

While there are many unanswered questions pertinent to brain function, necessitated by limited knowledge about brain, it is almost certain that the brain is a many body system which can accumulate and store energy. When needed, the stored energy can be released to accomplish missions. The communication between different parts in brain is likely mediated by certain brain waves, as no other means can convey information more effectively than waves. As long as these general discernments hold, given its generality the present model expects to play a role in guiding or stimulating further experimental and theoretical studies.

**Acknowledgments.** The work received support from National Science Council through the grants NSC89-2112-M008-008 and through Dr. H. C. Lee (NSC89-2816-M008-003-7).

\*This paper is based on a talk given at the Second Cross Taiwan Strait Biology Inspired Theoretical Studies (BITS) workshop, Beijing, May 18, 2000. The author thanks the organizers for kind hospitality.